\documentclass{ws-procs9x6}

\def\etal {{\it et al.}}

\begin{document}

\title{MACROSCOPIC OBJECTS, INTRINSIC SPIN, AND LORENTZ VIOLATION}

\author{DAVID W.\ ATKINSON,$^*$ MCCOY BECKER, and JAY D.\ TASSON}

\address{Physics and Astronomy Department, Carleton College\\
Northfield, MN 55057, USA\\
$^*$E-mail: atkinsod@carleton.edu}

\begin{abstract}
The framework of the Standard-Model Extension (SME) provides a relativistic quantum field theory for the study of Lorentz violation.  The classical, nonrelativistic equations of motion can be extracted as a limit that is useful in various scenarios.  In this work, we consider the effects of certain SME coefficients for Lorentz violation on the motion of macroscopic objects having net intrinsic spin in the classical, nonrelativistic limit.
\end{abstract}

\bodymatter

\section{Introduction}
The most successful description of physics today is provided by General Relativity and the Standard Model; however, the combined description is unacceptable
as one approaches the Planck scale.
The Standard-Model Extension (SME) seeks to provide relevant experimental guidance in addressing this issue by introducing a framework of all possible Lorentz-violating terms.\cite{lll}  The quest to experimentally detect or constrain the Lorentz-violating terms in the SME requires the ability to detect the impact of these terms at scales reasonable for experiment.

One way to search for relativity violations is to study the effects of these couplings on macroscopic objects at low energies.  The goal of this work is to search for the impact of relativity violating terms in the SME on the acceleration of macroscopic objects with intrinsic spin in the nonrelativistic limit.  This extends work investigating torques on such objects,\cite{bluhm} work considering spin-independent Lorentz-violating accelerations,\cite{bertschinger}
and work on the classical relativistic theory. \cite{russell}

The initial theoretical tool for our investigation is the Foldy-Wouthuysen transformation.\cite{foldy}  The motivation for this transformation is to extract the nonrelativistic hamiltonian from its relativistic counterpart.  The Foldy-Wouthuysen transformation is designed to reduce the off-diagonal portions of the Dirac hamiltonian in a systematic fashion, so that the nonrelativistic (top $2\times2$ block) component of the hamiltonian can be taken on its own to some degree of accuracy.

\section{Free hamiltonian and generic force}
In considering the nonrelativistic free-particle hamiltonian, we consider results proportional to two powers of momentum.\cite{lane} These take the form
\begin{equation}
H=\frac{p^2}{2m}+...-\frac{1}{m}C_{jk}p_jp_k+...\text{,}
\end{equation}
where 
\begin{eqnarray}
\label{bigc}
C_{jk}&=&c_{jk}+\tfrac{1}{2}c_{00}\delta_{jk}-\{[(d_{0j}+d_{j0})-\tfrac{1}{2m}(b_j+m d_{j0}+\tfrac{1}{2}m\epsilon_{jmn}g_{mn0}
\nonumber \\
&&
\hskip -12pt
+\tfrac{1}{2}\epsilon_{jmn}H_{mn})]\delta_{kl}+\tfrac{1}{2m}(b_l+\tfrac{1}{2}m\epsilon_{lmn}g_{mn0})\delta_{jk}-\epsilon_{jlm}(g_{m0k}+g_{mk0})\}\sigma^l
\nonumber\\
\end{eqnarray}
to first order in Lorentz violation. Using Hamilton's equations of motion, we can repackage the Lorentz-violating piece into an effective mass term $m_{jk}$, given by
\begin{equation}
m_{jk}=m\delta_{jk}+2mC_{(jk)}\text{,}
\end{equation}
where $C_{(jk)}$ denotes symmetrization on the indices $jk$;
$C_{(jk)}=\tfrac{1}{2}(C_{jk}+C_{kj})$. 
We produce a modified form of Newton's second law
\begin{equation}
\label{n2}
F_j=m_{jk}a_k\text{,}
\end{equation}
where the mass is now a matrix dependent on spin orientation with respect to the
coefficients for Lorentz violation.

\begin{figure}
\begin{center}
\psfig{file=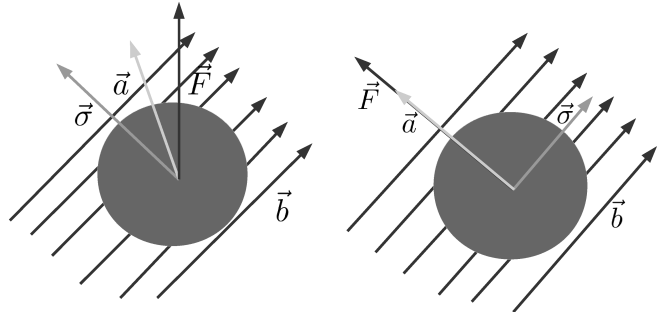,width=0.8\textwidth}
\end{center}
\caption{Two different cases of a particle with some intrinsic spin, $\vec{\sigma}$, in the presence of an arbitrary force $\vec{F}$ and the Lorentz-violating field $\vec{b}$.}
\label{figure}
\end{figure}

Figure \ref{figure} illustrates
two examples of motion
resulting from a nonzero $b_j$
in Eq.\ (\ref{bigc}). 
In the left-hand image, the particle has spin perpendicular to the direction of $\vec{b}$.  The net acceleration of this object is 
\begin{equation}
\vec{a} = \frac{\vec F}{m}-\frac{\vec{b}(\vec{\sigma}\cdot\vec{F})}{2 m^2}-\frac{\vec{\sigma}(\vec{b}\cdot\vec{F})}{2 m^2},
\end{equation}
with directionality affected by the orientation of $\vec{F}$ with respect to $\vec{b}$ and $\vec{\sigma}$.  The right-hand image shows the particle in the case of a force applied perpendicular to the direction of $\vec{b}$ and $\vec{\sigma}$.  Here the acceleration is
\begin{equation}
\vec a = \left(\frac{1}{m}+\frac{\vec{b}\cdot\vec\sigma}{m^2}\right)\vec{F}\text{.}
\end{equation}
Note that in this case,
the acceleration is in the usual direction,
but the magnitude is altered by $\vec{b}\cdot\vec\sigma$.

\section{Interactions}
In general,
the force appearing in Eq.\ (\ref{n2})
may contain additional Lorentz violation
and should be
handled by considering interactions in the hamiltonian. 
As an example,
we consider electromagnetism
with $b_\mu$
as the only nonzero coefficient for Lorentz violation.
We find a hamiltonian of the form
\begin{equation}
H = \frac{1}{2m}\left(\vec p - q \vec{A}\right)^2 -\frac{q}{2m} B_j \sigma^j + qA^0 -b_j\sigma^j + \frac{1}{m}(b_0p_j\sigma^j - qb_0 A_j\sigma^j),
\end{equation}
including terms to order in 1/m
in the Foldy-Wouthuysen expansion.  
To this order we find no additional $b_\mu$  contributions associated with the interaction, 
and the force appropriate for insertion into Eq.\ (\ref{n2}) takes the usual form:
\begin{equation}
F_j = qE_j + q\epsilon_{jkl}v_kB_l.
\end{equation}
A less trivial example
is provided by the case of gravitational
interactions.\cite{da}

\end{document}